\documentclass[aps,twocolumn,showpacs,pre]{revtex4-1}
\usepackage{graphicx,graphics}
\usepackage[colorlinks= true, linkcolor=blue, citecolor=blue, urlcolor=blue]{hyperref}
\usepackage{amsmath,amssymb,amsfonts}
\usepackage{bbm,bbold}
\usepackage{multirow}
\usepackage{hyperref}
\usepackage{graphicx}
\usepackage{color}
\usepackage{placeins}

\begin{document}

\title{Antifragility of stochastic transport on networks with damage}
\author{L. K. Eraso-Hernandez${}^1$, A. P. Riascos${}^{2}$}  
\affiliation{${}^1$Instituto de Física, Universidad Nacional Autónoma de México, Ciudad Universitaria, 04510, Mexico City, Mexico\\
${}^{2}$Departamento de F\'isica, Universidad Nacional de Colombia, Bogotá, Colombia}

\date{\today}

\begin{abstract}
A system is called antifragile when damage acts as a constructive element improving the performance of a global function. In this paper, we analyze the emergence of antifragility in the movement of random walkers on networks with modular structures or communities. The random walker hops considering the capacity of transport of each link, whereas the links are susceptible to random damage that accumulates over time. We show that in networks with communities and high modularity, the localization of damage in specific groups of nodes leads to a global antifragile response of the system improving the capacity of stochastic transport to more easily reach the nodes of a network. Our findings give evidence of the mechanisms behind antifragile response in complex systems and pave the way for their quantitative exploration in different fields.
\end{abstract}

\begin{titlepage}
\maketitle
\end{titlepage}
\section{Introduction}
\begin{figure}[b]
    \centering
    \includegraphics*[width=0.5\textwidth]{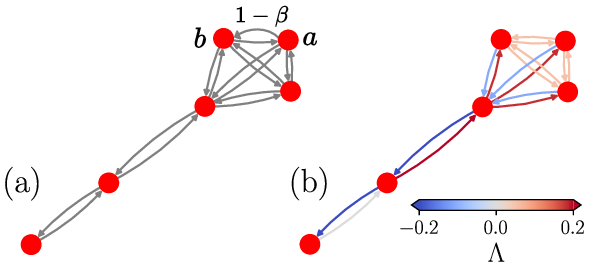}
    \vspace{-2mm}
    \caption{Global response to damage on a graph with six nodes. (a) Graph with a weight $1-\beta$ in the directed edge $(a,b)$. (b) Graph with edges colored using the values $\Lambda$ that quantify the response of the global transport to infinitesimal damage in each edge, $\Lambda>0$ evidences antifragility.  }
    \label{Fig_1}
\end{figure}

Complex systems are composed of many interconnected components and are highly sensitive
to perturbations \cite{BaryamBook_1997,SayamaBook}. These systems can be vulnerable to diverse types of failures exhibiting a reduction of their global functionality in response to damage 
\cite{VespiBook,Carlson_PRL2000,Carlson_PNAS_2002,Dobson2007,SunPNAS2020,VuralPRE2014,Cohen2016}. In this context, the study of dynamical processes in systems with damage accumulation is important to
identify vulnerabilities and
to better understand the relation between the dynamics and the structure of a system \cite{VespiBook,barabasi2016book}. Recent efforts include the study of optimal structures for transport \cite{Katifori_PRL2010}, network reparation after attacks \cite{Farr_PRL_2014,Hu2016,Lin2020}, the impact of damage in diffusive dynamics  \cite{Aging_PhysRevE2019,Eraso_Hernandez_2021,Eraso-metro} and synchronization of oscillators \cite{Eraso-Hernandez_2023} on networks, the response to targeted attacks or failure in interacting neuronal units \cite{FaciLazaro2022} and dynamic and functional alterations of rodent cortical networks
grown in vitro that were physically damaged \cite{TellerENEURO2020,mi13122259},  just to mention a few examples.
\\[2mm]
Some systems have the property of being robust \cite{Carlson_PRL2000}, in the sense that they are not significantly affected by the presence of damage. On the other extreme, systems may be fragile, where damage breaks or collapses the system.  In this context, a system is considered antifragile when it benefits from damage. Thus, damage acts as a constructive element that allows the system to perform a global function successfully. This idea is exemplified in Fig. \ref{Fig_1}. Figure  \ref{Fig_1}(a)  presents a network where a random walker hops using the information of weights in the edges, and each edge has a weight of $1$ with the exception of the directed edge $(a,b)$ with weight $1-\beta$, with $0\leq \beta < 1$. For $\beta=0$, the dynamics is defined by a standard random walk \cite{NohRieger2004,Hughes,MasudaPhysRep2017,ReviewJCN_2021} and the average number of steps (cover time) $t_{\mathrm{cover}}$ to start randomly from one node and visit at least once all the nodes is $t_{\mathrm{cover}}=28.2$. Counterintuitive, if damage increases to $\beta=0.5$, $t_{\mathrm{cover}}=27.4$ and, the complete removal of this link using $\beta\to 1$ produces $t_{\mathrm{cover}}=26.6$ (each  average cover time $t_{\mathrm{cover}}$ is obtained using $10^6$ realizations of the random walk dynamics). Then, the global capacity of the network to communicate all the nodes is improved with the removal of the link $(a,b)$. This property is not exclusive of the edge $(a,b)$; in Fig. \ref{Fig_1}(b) we show the effect of infinitesimal variations in the capacity of transport of each edge measured with $\Lambda$, if $\Lambda>0$ evidences antifragility, whereas  $\Lambda<0$ indicates the reduction in the global capacity of the random walker to explore the network [$\Lambda$ is defined in Eq. (\ref{Lambda_edge})].
\\[2mm]
The concept of antifragility was coined by Taleb and explained in his book \cite{Taleb_book}. Taleb’s ideas have been further explored in several contexts such as financial systems, stocks and cryptocurrencies \cite{Stocks_Alatorre2023}, the design of renewable energy systems \cite{Coppitters2023}, and ecosystems \cite{Equihua_PeerJ_2020}, among many others \cite{axenie2023antifragility}. Also, a similar phenomenon has been observed in simulations analyzing travel times of uncoordinated drivers on urban road networks, where blocking certain streets can partially improve global traffic conditions \cite{Hyejin_PRL2008}. However, despite its potential to understand the impact of damage in complex systems and potential applications in diverse fields, little is known about how to characterize antifragility and the nature of the systems in which it can emerge, and there is a need for precisely defining the scales at which antifragility operates \cite{axenie2023antifragility}.  Recent efforts include the modeling of antifragility in systems of Boolean networks \cite{Zanin2019,Lopez_Entropy_2023}, and the study of strongly coupled Coulomb fluids to have an understanding of the physics behind antifragile systems \cite{Ghodrat_2015}.
\\[2mm]
In this paper,  we analyze the antifragility of random walk dynamics on networks. The random walker hops considering the capacity of transport of each link, whereas the links are susceptible to damage that can affect networks with modular structures or communities. In Sec. \ref{Sec_Theory}, we present the formal definitions of the random walker considered and a global time that measures its performance to explore a network. In terms of this dynamic process, we introduce a general framework to detect and measure antifragility and the global response of a system to damage at the level of edges. In Sec. \ref{Sec_Results}, we discuss different examples of networks with damage in the links where the effects of antifragility are observed. We analyze lollipop graphs,  all the connected non-isomorphic graphs with sizes $N = 4$ and $N=5$, and the effect of damage that accumulates preferentially in specific parts of networks with cliques and communities. Finally, we explore antifragility in four real networks with communities. In Sec. \ref{Sec_Conclusions} we  present the conclusions. The methods introduced to characterize the global response to modifications in the capacity of transport associated with random walkers on networks are general and can be implemented to the study of other systems and processes throughout the definition of a global functionality and its response to changes in the interactions between the parts of the system. Our findings pave the way to a broad understanding of the emergence of antifragility and the impact of damage in complex systems.
\section{General theory}
\label{Sec_Theory}
\subsection{Global functionality and the effect of damage}
We consider a Markovian random walker on a connected undirected network with $N$ nodes $i=1,\ldots ,N$  defined by its adjacency matrix $\mathbf{A}$, with elements $A_{ij}=A_{ji}=1$ if the nodes $i$ and $j$ are connected to each other and 0 otherwise. The quantity $k_i=\sum_{l=1}^NA_{il}$ is the degree of the node $i$. In addition, we have a matrix of weights $\mathbf{\Omega}$ with elements $\Omega_{ij}\geq 0$. By definition, time is discrete, i.e., $t=0,1,2,\ldots$. The walker starts at $t=0$ from the node $i$ and at each step the random walker hops to a new node chosen with the transition probabilities in the matrix $\mathbf{W}(\mathbf{\Omega})$ with elements
\begin{equation}
	w_{i\to j}(\mathbf{\Omega})\equiv \frac{\Omega_{ij}}{\mathcal{S}_i},
\end{equation}
where $\mathcal{S}_i=\sum_{l=1}^N \Omega_{il}$ is the generalized strength of the node $i$ \cite{ReviewJCN_2021}.
The standard random walk strategy is defined using $\Omega_{ij}=A_{ij}$; in this manner, the walker moves from a node to one of its nearest neighbors with equal probability \cite{Lovasz1996,NohRieger2004,MasudaPhysRep2017}, and we denote as $\mathbf{W}$ (with elements $w_{i\to j}$) the transition matrix $\mathbf{W}(\mathbf{A})$.
\\[2mm]
The capacity of this process to explore the entire structure is measured by a global time $\tau(0)$ defined in terms of the eigenvectors and eigenvalues  of $\mathbf{W}$ \cite{ReviewJCN_2021}. Let us denote the eigenvalues of  $\mathbf{W}$ as $\lambda_\ell$ (where $\lambda_1=1$), and its right and left eigenvectors as $\left|\phi_\ell\right\rangle$ and $\left\langle\bar{\phi}_\ell\right|$, respectively, for $\ell=1,2,\ldots,N$. Then, $\mathbf{W}\left|\phi_\ell\right\rangle=\lambda_\ell\left|\phi_\ell\right\rangle $ and $\left\langle\bar{\phi}_\ell\right|\mathbf{W}=\lambda_\ell \left\langle\bar{\phi}_\ell\right|$ for $\ell=1,\ldots,N$, where the set of eigenvalues is ordered in the form $\lambda_1=1$ and $1>\lambda_\ell\geq -1$ for $\ell=2,3,\ldots,N$. We define $|i\rangle$ as the vector whose components are 0 except the $i$-th one, which is 1. In the following we denote as  $\{\left| i \right\rangle \}_{i= 1}^N$ the canonical base of $\mathbb{R}^N$ \cite{ReviewJCN_2021}. 
\\[2mm]
Using the information of the right eigenvectors, we define a matrix $\mathbf{Q}$ with elements $Q_{ij}=\left\langle i|\phi_j\right\rangle$. The matrix $\mathbf{Q}$ is invertible, and a new set of vectors $\left\langle \bar{\phi}_i\right|$ is obtained by $(\mathbf{Q}^{-1})_{ij}=\left\langle \bar{\phi}_i |j\right\rangle $, then
\begin{equation}\label{conditionOC}
	\delta_{ij}=(\mathbf{Q}^{-1}\mathbf{Q})_{ij}=\sum_{\ell=1}^N \left\langle\bar{\phi}_i|\ell\right\rangle \left\langle \ell|\phi_j\right\rangle=\langle\bar{\phi}_i|\phi_j\rangle \, ,
\end{equation}
and
\begin{equation}\label{conditionC}
	\mathbf{I}=\mathbf{Q}\mathbf{Q}^{-1}=\sum_{\ell=1}^N \left|\phi_\ell\right\rangle \left\langle \bar{\phi}_\ell \right| \, ,
\end{equation}
where $\mathbf{I}$ is the $N\times N$ identity matrix. In addition, by normalization of the probability, the matrix $\mathbf{W}$ is such that $\sum_{\ell=1}^N w_{i \to \ell}=1$, which implies that $|\phi_1 \rangle\propto \left(\begin{array}{c} 1\\ 1 \\ \ldots \\ 1\end{array}\right)$. 
\\[2mm]
In terms of the eigenvalues and eigenvectors of $\mathbf{W}$, the global time $\tau(0)$ is given by \cite{ReviewJCN_2021}
\begin{equation}\label{tau_global0}
\tau(0)\equiv\frac{1}{N}\sum_{j=1}^N \tau_j(0),
\end{equation}
with \cite{ReviewJCN_2021}
\begin{equation}\label{tau_j0}
	\tau_j(0)=\frac{1}{P_j^\infty}\sum_{\ell=2}^N \frac{1}{1-\lambda_\ell}Z_{jj}^{(\ell)},
\end{equation}
where $Z_{jj}^{(\ell)}\equiv\left\langle j|\phi_\ell\right\rangle \left\langle\bar{\phi}_\ell|j\right\rangle$ and $P_j^\infty= \left\langle j|\phi_1\right\rangle \left\langle\bar{\phi}_1|j\right\rangle$ is the stationary probability distribution of the random walker.
\\[2mm]
Let us now define a modified dynamics with a reorganization of the transition probabilities in $\mathbf{W}$ due to a reduction of the capacity of transport (damage) in the link $(a,b)$ connecting $a$ to $b$ in the network. This modification is defined by the matrix of weights $\mathbf{\Omega}^\star$ with elements $\Omega^\star_{ab}=(1-\beta)A_{ab}$ and $\Omega^\star_{ij}=A_{ij}$ in other cases.
The transition matrix $\mathbf{W}(\mathbf{\Omega}^\star)$ defines an ergodic process for  $0\leq \beta<1$. In particular, due to the normalization, $\mathbf{W}(\mathbf{\Omega}^\star)$, and the standard random walk defined by $\mathbf{W}$ differ in the $a$-th row. Also, it is possible, at least numerically, to obtain a global time denoted as $\tau(\beta)$ similar to $\tau(0)$, but now using the eigenvalues and eigenvectors of  $\mathbf{W}(\mathbf{\Omega}^\star)$. In terms of these quantities, we define a global functionality \cite{Aging_PhysRevE2019,Eraso_Hernandez_2021}
\begin{equation}\label{F_beta_edge}
\mathcal{F}_\beta\equiv\frac{\tau(0)}{\tau(\beta)},
\end{equation}
that measures the global response of the system  to the modification in the edge $(a,b)$. In particular, if we consider the effect of an infinitesimal reduction in the capacity of the link, it is convenient to define
\begin{equation}\label{Lambda_edge}
\Lambda\equiv\frac{d\mathcal{F}_\beta}{d\beta}\Big|_{\beta\to 0}.
\end{equation}
\\
Then, $\Lambda>0$ evidences antifragility, whereas $\Lambda<0$ is associated with the reduction of the capacity of transport with the infinitesimal damage in the link.  $\Lambda=0$ is obtained in cases where the infinitesimal variation of the weight in a specific link does not modify 
$\tau(\beta)$.
\subsection{Perturbation approach for the calculation of $\Lambda$}\label{S0}
In this section, we study the effect of infinitesimal modifications in the weight of an edge using perturbation theory to deduce an analytical expression for $\Lambda$. We consider a modified dynamics defined by a random walker with a transition matrix $\mathbf{\Pi}^{(a,b)}_\beta$ that requires an adjustment in the elements of $\mathbf{W}$ due to a reduction of the capacity of transport (damage) of the link $(a,b)$ connecting $a$ to $b$ in the network. This modification is defined by the matrix of weights $\mathbf{\Omega}^\star$ with elements $\Omega^\star_{ab}=(1-\beta)A_{ab}$ with $0\leq \beta<1$ and $\Omega^\star_{ij}=A_{ij}$ in other cases. Then, the elements of $\mathbf{\Pi}^{(a,b)}_\beta=\mathbf{W}(\mathbf{\Omega}^\star)$ take the form
\begin{align}
	\left(\mathbf{\Pi}^{(a,b)}_\beta\right)_{ij}&\equiv  \frac{\Omega^\star_{ij}}{\sum_{l=1}^N \Omega^\star_{il}},\\
	&=
	\begin{cases}
		\displaystyle
		w_{i\to j}\qquad &\mathrm{if}\qquad i\neq a,\\
		\displaystyle
		\frac{A_{aj}}{k_a-\beta} &\mathrm{if}\qquad i=a,\,\, j\neq b,\\
		\displaystyle
		\frac{(1-\beta)A_{ab}}{k_a-\beta} &\mathrm{if}\qquad i=a,\,\, j=b.\\
	\end{cases}
\end{align}
Therefore, for $\beta\ll 1$, the transition matrix $\mathbf{\Pi}^{(a,b)}_\beta$ for the dynamics with a small reduction $\beta$ in the weight of the link $(a,b)$ is
\begin{equation}\label{Pi_def}
	\mathbf{\Pi}^{(a,b)}_\beta\approx\mathbf{W}+\beta\left(\frac{\mathbf{E}^{(a,a)}\mathbf{W}-\mathbf{E}^{(a,b)}}{k_a}\right),
\end{equation}
where the matrix  $\mathbf{E}^{(a,b)}$ is associated with the modification of the edge $(a,b)$ with elements $\left(\mathbf{E}^{(a,b)}\right)_{ij}\equiv \delta_{ia}\delta_{jb}$. In this manner, defining
\begin{equation}\label{Gamma_matrix_def}
	\mathbf{\Gamma}\equiv\frac{\mathbf{E}^{(a,a)}\mathbf{W}-\mathbf{E}^{(a,b)}}{k_a},
\end{equation}
we have for Eq. (\ref{Pi_def})
\begin{equation}\label{Pi_def_Gamma}
	\mathbf{\Pi}^{(a,b)}_\beta\approx\mathbf{W}+\beta\mathbf{\Gamma}.
\end{equation}
Here, it is important to notice that for $0\leq\beta<1$, $\mathbf{\Pi}^{(a,b)}_\beta$ in Eq. (\ref{Pi_def_Gamma}) is a stochastic matrix since $\sum_{l=1}^N (\mathbf{\Gamma})_{il}=0$.
\\[2mm]
In the following, we are interested in the effect of small $\beta$ in the functionality $\mathcal{F_\beta}$ that characterizes global transport in Eq. (\ref{F_beta_edge}).  In this case, the value $\tau(\beta)$ is obtained using in Eqs. (\ref{tau_global0}) and (\ref{tau_j0}) the eigenvalues and eigenvectors of $	\mathbf{\Pi}^{(a,b)}_\beta$. In terms of the functionality $\mathcal{F_\beta}$ and $\Lambda$ in Eq. (\ref{Lambda_edge}), we have
\begin{equation}
\label{Lambda_def}
	\Lambda=-\frac{\tau(0)}{(\tau(\beta))^2}\frac{d\tau(\beta)}{d\beta}\Big|_{\beta\to 0}=-\frac{1}{\tau(0)}\frac{d\tau(\beta)}{d\beta}\Big|_{\beta\to 0}.
\end{equation}
In addition, for small $\beta$
\begin{equation}\label{tau_beta_Delta}
	\tau(\beta)=\tau(0)+\beta\Delta\tau+o(\beta^2),
\end{equation}
where $\Delta\tau$ captures at first order the effect of the damage in $\tau(\beta)$. Therefore, Eq. (\ref{Lambda_def}) takes the form
\begin{equation}\label{Lambda_Delta_tau}
	\Lambda=-\frac{\Delta \tau}{\tau(0)}.
\end{equation}
The introduction of damage modifies the eigenvalues and eigenvectors of the original matrix $\mathbf{W}$. Expressing the modified values of $\tau_j(0)$ for small  $\beta$ as
\begin{equation}\label{tau_jbeta}
	\tau_j(\beta)\approx\frac{1}{P_j^\infty+\beta \Delta P_j^\infty}\sum_{\ell=2}^N \frac{Z_{jj}^{(\ell)}+\beta\Delta Z_{jj}^{(\ell)}}{1-\lambda_\ell-\beta \Delta\lambda_\ell} ,
\end{equation}
where $\Delta\lambda_\ell$, $\Delta P_j^\infty$ and $\Delta Z_{jj}^{(\ell)}$ denote the first order corrections for the eigenvalues, stationary probability distribution and  $Z_{jj}^{(\ell)}$ associated to $\mathbf{\Pi}^{(a,b)}_\beta$. However, using Eq. (\ref{tau_jbeta}) for $\beta\ll 1$
\begin{align}\nonumber
	\tau_j(\beta)&\approx\frac{1}{P_j^\infty}\left(1-\beta\frac{\Delta P_j^\infty}{P_j^\infty}\right)\\ \nonumber
	&\times\sum_{\ell=2}^N \frac{1}{1-\lambda_\ell}\left(1+\beta \frac{\Delta\lambda_\ell}{1-\lambda_\ell}\right)(Z_{jj}^{(\ell)}+\beta\Delta Z_{jj}^{(\ell)})\\
	&= \tau_j(0)+\beta \Delta \tau_j+o(\beta^2),
\end{align}	
with 
\begin{multline}\label{delta_tau_jbeta}
	\Delta \tau_j=\frac{1}{P_j^\infty}\sum_{\ell=2}^N \frac{1}{1-\lambda_\ell}\Delta Z_{jj}^{(\ell)}+\frac{1}{P_j^\infty}\sum_{\ell=2}^N \frac{\Delta\lambda_\ell}{(1-\lambda_\ell)^2}Z_{jj}^{(\ell)}\\-\frac{\Delta P_j^\infty}{(P_j^\infty)^2}\sum_{\ell=2}^N \frac{1}{1-\lambda_\ell}Z_{jj}^{(\ell)}.
\end{multline}
As a consequence, from the definition in Eq. (\ref{tau_beta_Delta}), we obtain
\begin{equation}\label{Delta_tau}
	\Delta \tau=\frac{1}{N}\sum_{j=1}^N \Delta \tau_j=\mathcal{T}_1+\mathcal{T}_2-\mathcal{T}_3,
\end{equation}
with
\begin{align}\label{Time_1}
	\mathcal{T}_1&\equiv\sum_{\ell=2}^N \frac{1}{1-\lambda_\ell}\left(\sum_{j=1}^N  \frac{\Delta Z_{jj}^{(\ell)}}{N P^\infty_j} \right),\\
	\label{Time_2}
	\mathcal{T}_2&\equiv\sum_{\ell=2}^N \frac{\Delta\lambda_\ell}{(1-\lambda_\ell)^2}\left(\sum_{j=1}^N  \frac{Z_{jj}^{(\ell)}}{N P^\infty_j} \right),\\
	\label{Time_3}
	\mathcal{T}_3&\equiv\frac{1}{N}\sum_{j=1}^N \frac{\Delta P^\infty_j}{P^\infty_j}\tau_j(0).
\end{align}	
In this manner, using Eq. (\ref{Lambda_Delta_tau})
\begin{equation}\label{Lambda_def_Ts}
	\Lambda=-\frac{1}{\tau(0)}\left( \mathcal{T}_1+\mathcal{T}_2-\mathcal{T}_3\right).
\end{equation}
The result for $\Lambda$ in Eq. (\ref{Lambda_def_Ts}) can be obtained analytically using perturbation theory to determine the values $\mathcal{T}_1$, $\mathcal{T}_2$, and $\mathcal{T}_3$ in terms of the eigenvalues and eigenvectors of the transition matrix $\mathbf{W}$. This approach allows obtaining for the case of nondegenerate eigenvalues (see Appendix \ref{Appendix_Lambda} for details)
\begin{equation}\label{T_1eigs}
	\mathcal{T}_1=\sum_{\ell=2}^N \frac{1}{1-\lambda_\ell}X^{(\ell)},
\end{equation}
with
\begin{multline}
	X^{(\ell)}=\sum_{j=1}^N  \frac{1}{N P^\infty_j} \sum_{m\neq \ell}\frac{\left\langle j|\phi_m\right\rangle \left\langle\bar{\phi}_\ell|j\right\rangle\langle\bar{\phi}_m| \mathbf{\Gamma} |\phi_\ell\rangle}{\lambda_\ell-\lambda_m}
	\\
	+
	\sum_{j=1}^N  \frac{1}{N P^\infty_j}\sum_{m\neq \ell}
	\frac{
		\left\langle j|\phi_\ell\right\rangle \left\langle\bar{\phi}_m|j\right\rangle\langle\bar{\phi}_\ell| \mathbf{\Gamma} |\phi_m\rangle}{\lambda_\ell-\lambda_m},
\end{multline}
for $\ell=2,3,\ldots,N$. Similarly, 
\begin{align}
	\mathcal{T}_2&=\sum_{\ell=2}^N \frac{\langle\bar{\phi}_\ell| \mathbf{\Gamma} |\phi_\ell\rangle}{(1-\lambda_\ell)^2}\left(\sum_{j=1}^N  \frac{Z_{jj}^{(\ell)}}{N P^\infty_j} \right),\\
	\mathcal{T}_3&=\sum_{j=1}^N \frac{\tau_j(0)}{N P^\infty_j}\left(\sum_{m=2}^N\frac{ \left\langle\bar{\phi}_m|j\right\rangle \left\langle j|\phi_1\right\rangle \langle\bar{\phi}_1| \mathbf{\Gamma} |\phi_m\rangle}{1-\lambda_m}\right). \label{T_3eigs}
\end{align}	
In this manner, the analytical results in Eqs. (\ref{T_1eigs})-(\ref{T_3eigs}) show that from the information of eigenvalues and eigenvectors of the transition matrix $\mathbf{W}$, we can measure the effect of infinitesimal damage on the edge associated with the matrix $\mathbf{\Gamma}$ in Eq. (\ref{Gamma_matrix_def}). The combination of values $\mathcal{T}_1$, $\mathcal{T}_2$, and $\mathcal{T}_3$ in $\Lambda$ given by Eq. (\ref{Lambda_def_Ts})  allows us to determine whether the damage produces an antifragile response (when $\Lambda>0$). The results are valid only when the eigenvalues are nondegenerate; however, the approach explored shows the effect that $\mathbf{\Gamma}$  has on $\Lambda$. The impact of damage in general cases can be obtained numerically using the eigenvalues and eigenvectors of $\mathbf{W}$ and  $\mathbf{\Pi}^{(a,b)}_\beta=\mathbf{W}(\mathbf{\Omega}^\star)$ to evaluate directly Eq. (\ref{Lambda_edge}).

\section{Results}
\label{Sec_Results}
\subsection{Lollipop graphs}
\label{Sec_Lollipop}
\begin{figure}[b!]
	\centering
	\includegraphics*[width=0.45\textwidth]{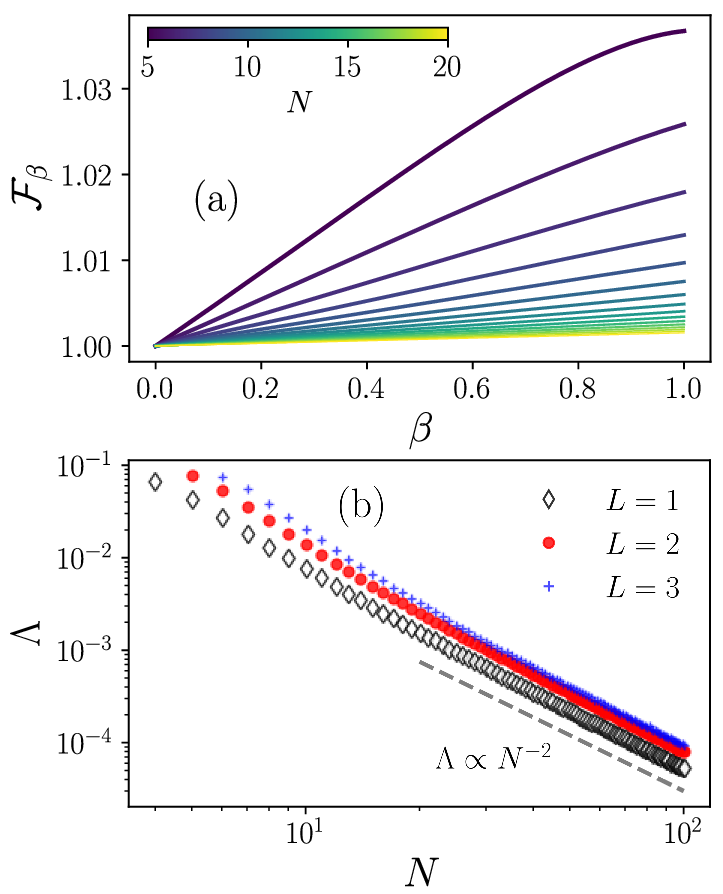}
	\vspace{-2mm}
	\caption{Antifragility in lollipop graphs with size $N$. (a) $\mathcal{F}_\beta$ in Eq. (\ref{F_beta_edge}) as a function of $\beta$ for lollipops with $L=1$, the colorbar codifies the sizes $N$ explored. (b) $\Lambda$ in Eq. (\ref{Lambda_edge}) in terms of the size $N$ for  graphs with $L=1,\,2,\,3$. The dashed line shows the inverse power-law relation $\Lambda\propto N^{-2}$.}
	\label{Fig_2}
\end{figure}
\begin{figure*}[t!]
	\centering
	\includegraphics*[width=0.97\textwidth]{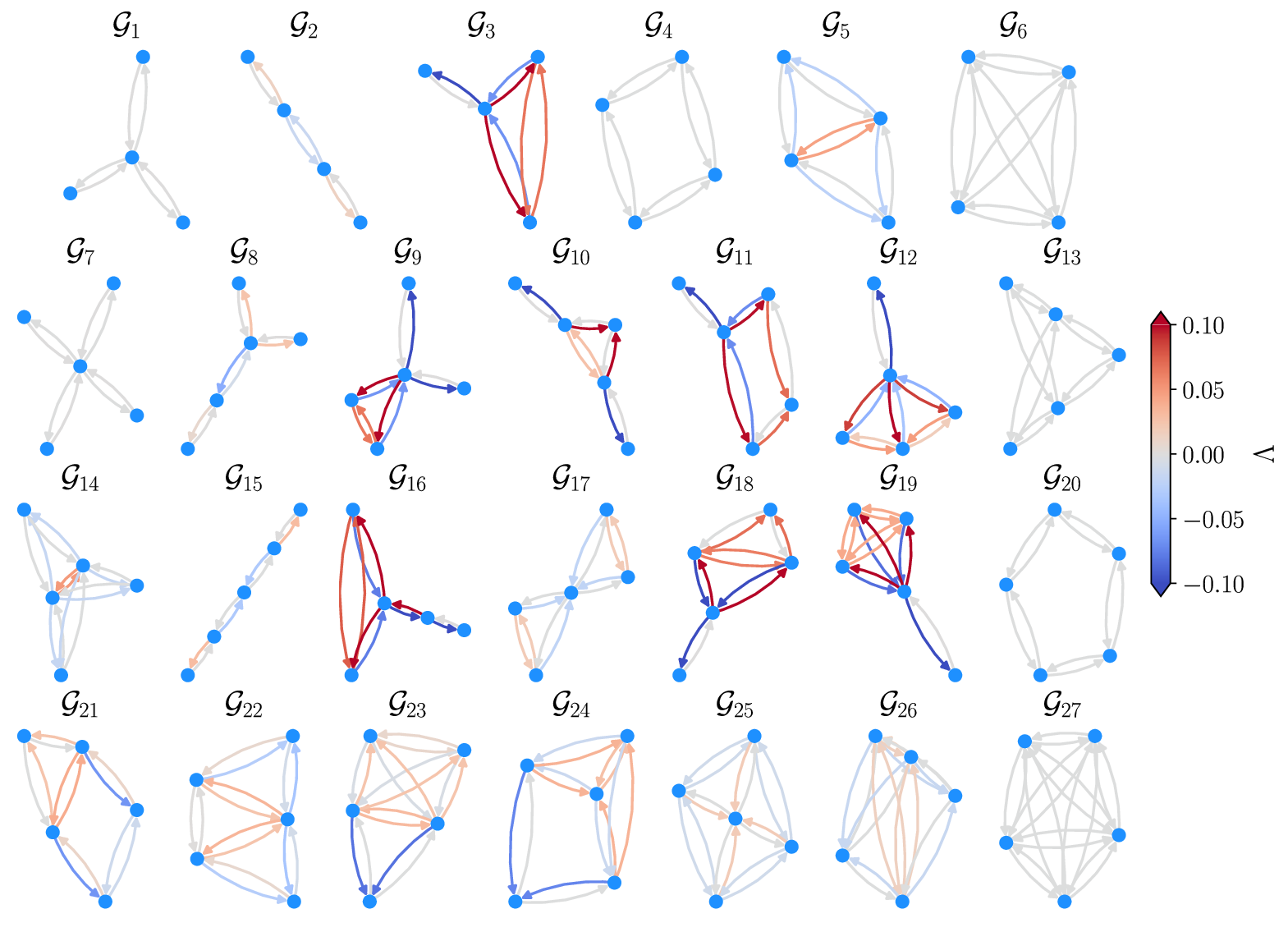}
	\vspace{-2mm}
	\caption{$\Lambda$ for all the nonisomorphic connected graphs with $N=4$ ($\mathcal{G}_1,\ldots,\mathcal{G}_6$) and $N=5$ ($\mathcal{G}_7,\ldots,\mathcal{G}_{27}$)  \cite{ConnectedGraphs}. The values are obtained numerically using Eq. (\ref{Lambda_edge}) considering infinitesimal damage in each link colored with $\Lambda$ encoded in the colorbar.  }
	\vspace{2mm}
	\label{Fig_3}
\end{figure*}
In this section, we calculate the global functionality $\mathcal{F}_\beta$ in Eq. (\ref{F_beta_edge}) and $\Lambda$ in Eq. (\ref{Lambda_edge}) for a set of graphs called lollipops in cases where the alteration of a particular link produces an antifragile response at a global level. A lollipop of size $N$ is a special type of network formed by a fully connected graph (clique) with $N-L$ vertices and a path graph (linear graph) with $L$ nodes. Both subgraphs are connected with an additional link called a bridge. In Fig.  \ref{Fig_1}(a), we show a lollipop with $N=6$ and $L=2$. In Fig. \ref{Fig_2} we show $\mathcal{F}_\beta$ and $\Lambda$ in lollipop graphs with different sizes $N$ and $L=1,2,3$. We analyze the effect of reducing the weight of a link $(a,b)$ in the clique that is not connected directly with the node that forms part of the bridge [this type of edge is illustrated in the link $(a,b)$ in Fig. \ref{Fig_1}(a)]. We center our discussion on this special link since its modification improves the global capacity of transport on the network.
\\[2mm]
In Fig. \ref{Fig_2}(a) we present the numerical results for   $\mathcal{F}_\beta$ as a function of $\beta$ with $0\leq \beta <1$ for graphs with $L=1$ and sizes $5\leq N\leq 20$ codified in the colorbar. In all the cases,  $\mathcal{F}_\beta>1$ for $0<\beta<1$ show that the capacity of transport of the network is increased with the damage $\beta$ in the link $(a,b)$.  The numerical values also reveal that the effect of the reduction of the weight in this link has a larger overall impact on structures with small sizes.
\\[2mm]
Similarly, the antifragility observed in Fig. \ref{Fig_2}(a) can be explored by considering the information of the derivative of $\mathcal{F}_\beta$ evaluated at $\beta=0$. In this manner, the slope of the tangent line of $\mathcal{F}_\beta$ at $\beta=0$ gives $\Lambda$ allowing to observe when a tiny infinitesimal link modification increases the functionality. In Fig. \ref{Fig_2}(b), we present $\Lambda$ as a function of $N$ for lollipops with $L=1,2,3$ and sizes $4\leq N \leq 100$. The results are obtained numerically using the approximation of Eq. (\ref{Lambda_edge}): $\Lambda\approx (\mathcal{F}_{\Delta \beta}-1)/\Delta \beta$ with $\Delta \beta=10^{-4}$. In all the cases analysed, $\Lambda>0$, evidencing the antifragility of this part of the structure. Also, for the cases explored $\Lambda\sim N^{-2}$ for $N\gg 1$.
\subsection{Response to damage in small graphs}
The method implemented in Sec. \ref{Sec_Lollipop} to study the antifragility in lollipops can be used to analyze the effect of infinitesimal damage in each link in a graph to detect parts that benefit, maintain, or reduce global functionality. In this section, we explore $\Lambda$ for the edges in all the connected nonisomorphic graphs with sizes $N = 4$ and $N=5$. The set of graphs analyzed is available in Ref. \cite{ConnectedGraphs} and contains 27 graphs that provide a great variety of topologies. In Fig. \ref{Fig_3}, we present the numerical results for $\Lambda$ evaluating the damage in each link for all the connected nonisomorphic graphs $\mathcal{G}_1,\ldots,\mathcal{G}_6$  with $N=4$ and $\mathcal{G}_7,\ldots,\mathcal{G}_{27}$ with $N=5$. In this representation, each link is colored with the respective $\Lambda$ obtained numerically using Eq. (\ref{Lambda_edge}), the values are codified in the colorbar.  A similar analysis is presented in Fig. \ref{Fig_1}(b) for a lollipop with $N=6$.
\\[2mm]
The results depicted in Fig. \ref{Fig_3} show that for the graphs explored the antifragility is not present in regular networks (structures with the same degree in each node) like the rings $\mathcal{G}_4$ and $\mathcal{G}_{20}$ and the fully connected graphs  $\mathcal{G}_6$,  $\mathcal{G}_{27}$ for which $\Lambda\to 0$ in all the edges. Something similar is observed in structures with high symmetry around a central node like in the star graphs $\mathcal{G}_1$ and $\mathcal{G}_{7}$. 
\\[2mm]
Our findings show that an important condition for antifragility to occur is to have a heterogeneous structure with groups of nodes with higher connectivity and other nodes with lower density of edges. This is the case of the lollipops $\mathcal{G}_3$,  $\mathcal{G}_{16}$,  $\mathcal{G}_{19}$, graphs with cliques like  $\mathcal{G}_{9}$,  $\mathcal{G}_{10}$, or other graphs similar to lollipops but with the clique replaced by a structure with some edges removed (see, for example, $\mathcal{G}_{11}$,  $\mathcal{G}_{12}$,  $\mathcal{G}_{18}$). The sufficient heterogeneity in the structure for the emergence of antifragility is also present in the tree graphs $\mathcal{G}_2$, $\mathcal{G}_8$, $\mathcal{G}_{15}$ and other networks like $\mathcal{G}_{5}$, $\mathcal{G}_{14}$,  $\mathcal{G}_{17}$ (formed by two cliques sharing a node) and $\mathcal{G}_{21},\ldots,\mathcal{G}_{26}$. Finally, it is important to notice that in the graphs analyzed, in structures with links with $\Lambda>0$ are also found parts of the network with $\Lambda<0$ showing that the antifragility in a complex system requires the existence of fragile parts.

\subsection{Damage accumulation in networks with communities}
\begin{figure*}[t!]
	\centering
	\includegraphics*[width=1.0\textwidth]{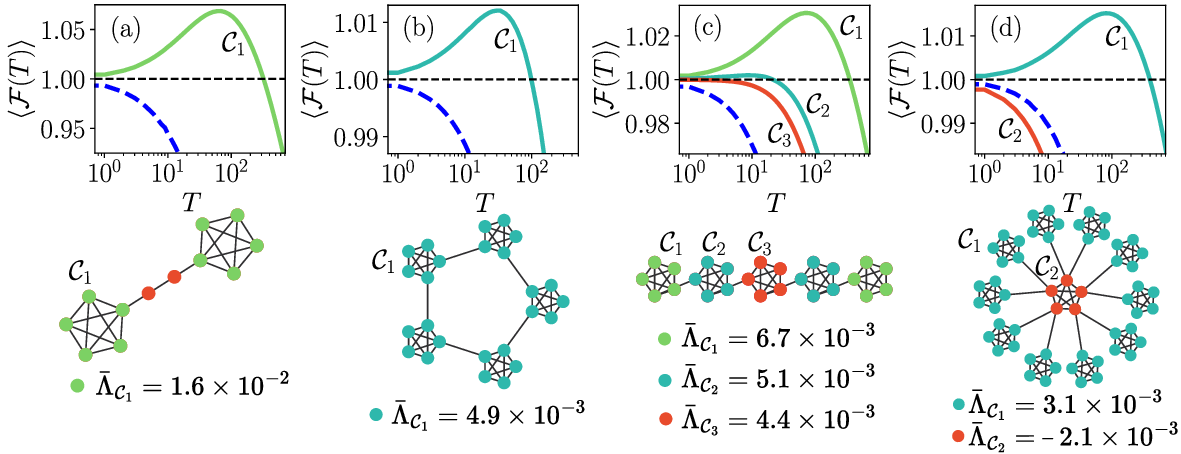}
	\vspace{-6mm}
	\caption{Antifragility in graphs with cliques. Solid lines depict $\langle \mathcal{F}(T)\rangle$ as a function of $T$ when damage affects communities in: (a) a barbell graph, (b) a chain of cliques, (c) a ring of cliques, and (d) a star of cliques. The thick dashed line corresponds to $\langle \mathcal{F}(T)\rangle$ when the damage affects the entire network. Ensemble averages were calculated using $10^5$ simulations with $\alpha=0.5$.}
	\label{Fig_4}
\end{figure*}
The previous analysis can be conducted on a different scale, where damage extends beyond a single link. This is important because complex systems are not only subject to localized damage but  also exposed to failure in specific areas. In the following, we explore the presence of antifragility in networks, where damage spreads to groups of edges, particularly within communities. A community of the network $\mathcal{G}$ with a set of edges $\mathcal{E}$ is a subgraph $\mathcal{C}$  defined as a subset of nodes that are densely connected to each other but share relatively few edges with other parts of the network \cite{Girvan-Newman2002,Newman2006}. As an extension of $\Lambda$, for a community $\mathcal{C}_i$, we define
\begin{equation}\label{Lambda_Community}
	\bar{\Lambda}_{\mathcal{C}_i}\equiv\frac{1}{|\mathcal{E}_i|}\sum_{(a,b)\in \mathcal{E}_i}\Lambda_{(a,b)},
\end{equation}
where $\mathcal{E}_i\subset\mathcal{E}$ is the set of edges within $\mathcal{C}_i$, $|\mathcal{E}_i|$ is the number of edges in this community and $\Lambda_{(a,b)}$ is obtained using Eq. (\ref{Lambda_edge}) for the edge $(a,b)$. Then, $\bar{\Lambda}_{\mathcal{C}_i}$ evaluates the effect of damage in $\mathcal{C}_i$,  $\bar{\Lambda}_{\mathcal{C}_i}>0$ indicates antifragility, whereas $\bar{\Lambda}_{\mathcal{C}_i}<0$ evidences fragility.
\\[2mm]
Now, to test the measure in Eq. (\ref{Lambda_Community}), let us describe the process of generating stochastic  cumulative damage in networks introduced in Refs. \cite{Aging_PhysRevE2019, Eraso_Hernandez_2021}.  This method employs a preferential attachment mechanism to reduce the functionality of edges in a network and has been implemented to study the effects of damage in  diffusive transport \cite{Aging_PhysRevE2019, Eraso_Hernandez_2021}, the analysis of infrastructure of metro systems \cite{Eraso-metro} and synchronization processes \cite{Eraso-Hernandez_2023}. In this framework,  damage occurs over a  temporal scale characterized by time $T$, the value of $T$ increases by 1 each time the network receives a failure on any of its edges. At time $T$ the probability that the edge $(i,j)$ receives damage is \cite{Aging_PhysRevE2019, Eraso_Hernandez_2021}
\begin{equation}\label{problinks}
\pi_{ij}(T)=\frac{h_{ij}(T-1)}{\sum_{(l,m) \in \mathcal{E}} h_{lm}(T-1)},
\end{equation}
where  $h_{ij}(T)$ is an integer variable. For $T=0$, $h_{ij}(0)=1$,  so $h_{ij}(T)-1$ gives us the number of accumulated faults  on  edge $(i,j)$ at time $T$. In this scenario, a matrix of weights   $\mathbf{\Omega}(T)$ describes the state of damage of the network. Its elements are defined by \cite{Aging_PhysRevE2019, Eraso_Hernandez_2021}
\begin{equation}
\Omega_{ij}(T)\equiv h_{ij}(T)^{-\alpha}A_{ij},
\end{equation}
where  $\alpha\geq0$ is a parameter of the model. If $\alpha=0$, the network can be repaired perfectly, preventing the accumulation of damage. Conversely, if $\alpha\to\infty$, network irreparability results in edge removal. 
\\[2mm]
An important feature of the probabilities in Eq. (\ref{problinks}) is that they produce preferential damage if a link has already suffered damage in the past. A link has a higher probability to get a fault with respect to a link never being damaged. Such preferential random processes have been explored in different contexts in science (see Ref. \cite{barabasi2016book}), being a key element in our model that generates complexity in the distribution of damage reflected by asymptotically emerging power-law and fractal features. An asymptotic analysis of the time evolution of the fault number distribution resulting from Eq. (\ref{problinks})
shows that a power-law scaling with features of a stochastic fractal emerges (see Ref. \cite{Aging_PhysRevE2019}). Such a preferential 
damage accumulation mechanism can be observed in several adaptive complex systems, such as living beings, and was suggested as a model for aging  \cite{Aging_PhysRevE2019}.
\\[2mm]
In addition, in the context of transport on networks, we use the  functionality $\mathcal{F}(T)$ that quantifies the global transport capacity at time $T$ as \cite{Aging_PhysRevE2019}
\begin{equation}
\mathcal{F}(T)\equiv\frac{\tau(0)}{\tau(T)},
\end{equation}
where $\tau(T)$ is the global time of the random walker defined by the transition matrix $\mathbf{W}(\mathbf{\Omega}(T))$. Since damage occurs stochastically, we denote as  $\langle\mathcal{F}(T)\rangle$   the ensemble average of  functionality obtained for different realizations.
\subsubsection{Antifragility in networks with cliques}
In the following, we implement the algorithm of cumulative damage in networks with fully connected communities represented by cliques. In Fig. \ref{Fig_4}, the panels present with solid lines $\langle\mathcal{F}(T)\rangle$  when the damage is localized in different communities (cliques) $\mathcal{C}_i$, along with the respective values $\bar{\Lambda}_{\mathcal{C}_i}$ obtained using Eq. (\ref{Lambda_Community}). The thick dashed line in each panel depicts $\langle\mathcal{F}(T)\rangle$ when the damage affects the entire structure. Ensemble averages are calculated using $10^5$ realizations of Monte Carlo simulations with $\alpha=0.5$.
\\[2mm]
In Fig. \ref{Fig_4}(a), we study the effect of damage in a barbell graph. This structure has two cliques $\mathcal{C}_1$  joined by a linear graph. Using Eq. (\ref{Lambda_Community}), we found $\bar{\Lambda}_{\mathcal{C}_1}=1.6\times10^{-2}$, a positive value that aligns with the results of $\langle\mathcal{F}(T)\rangle$ that exhibits antifragility. In particular, it is observed that when the edges in one of the cliques receive damage,  $\langle\mathcal{F}(T)\rangle$ takes values greater than 1 for different $T$. This result suggests that, on average, the global time of the random walker decreases as a consequence of the damage. In contrast, when the entire network is affected by damage, $\langle\mathcal{F}(T)\rangle$ only takes values less than 1. It is worth noticing that after a considerable amount of damage to the clique, the global transport is affected, as indicated by $\langle\mathcal{F}(T)\rangle<1$ .
\\[2mm]
\begin{figure*}[t!]
	\centering
	\includegraphics*[width=1.0\textwidth]{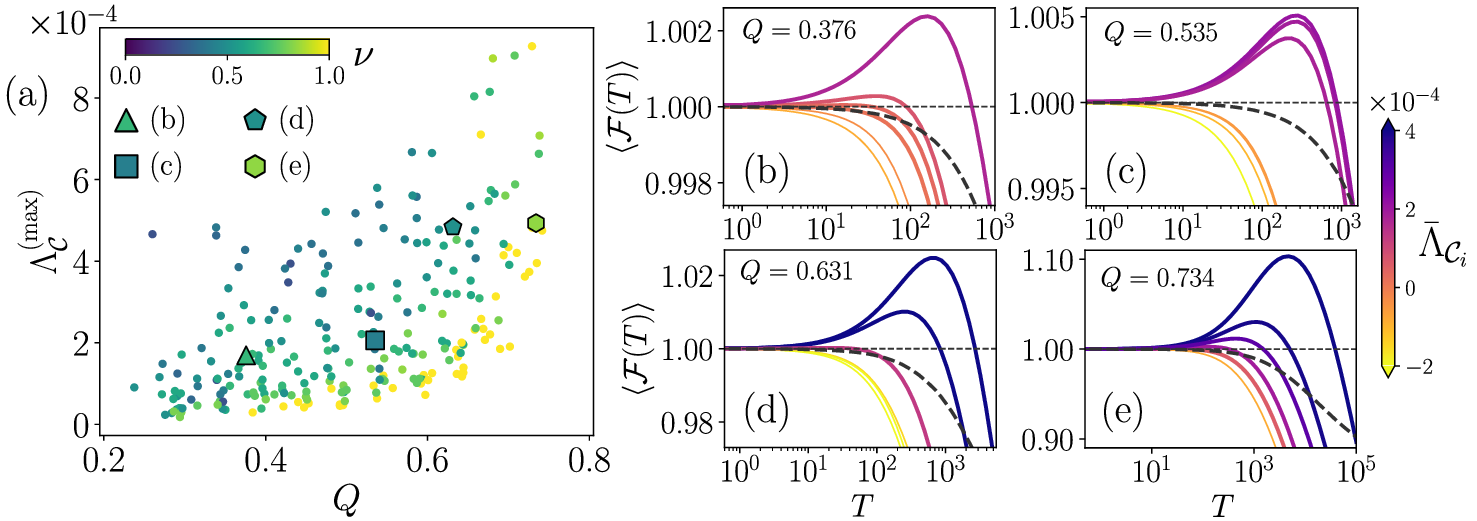}
	\vspace{-4mm}
	\caption{Effect of damage on networks with communities with $N=100$. (a) $\Lambda^{(\mathrm{max})}_\mathcal{C}$ as a function of $Q$. The fraction of antifragile communities  $\nu$  is encoded in the colorbar. (b) $\langle\mathcal{F}(T)\rangle$ as a function of $T$ for a network  with $Q=0.376$ when damage affects its communities.  The thick dashed line corresponds to $\langle\mathcal{F}(T)\rangle$ when damage is distributed across the entire network, $\bar{\Lambda}_{\mathcal{C}_i}$ for the communities is codified by the colorbar. Panels (c)-(e) correspond to the analysis of networks with different values $Q$. The values $\langle\mathcal{F}(T)\rangle$  are calculated using $10^4$ Monte Carlo simulations and $\alpha=0.5$. }
	\label{Fig_5}
\end{figure*}
In  Fig. \ref{Fig_4}(b), we present a ring structure, composed of five cliques connected by single edges. In this case, if any of the cliques in the ring receives damage, the network responds with antifragility. The values of  $\langle\mathcal{F}(T)\rangle$ show an improvement in the transport capacity of the network for different $T$, which is also indicated by $\bar{\Lambda}_{\mathcal{C}_1}=4.9\times10^{-3}$. However, when damage spreads across the entire structure, we observe only a degradation in the global functionality of the network.  
\\[2mm]
Similarly, for the chain formed by five cliques depicted in  Fig. \ref{Fig_4}(c), the topology of this structure leads to different responses of the network depending on which clique accumulates damage. The results show the most antifragile response in the cliques $\mathcal{C}_1$ at the end of the chain, with $\bar{\Lambda}_{\mathcal{C}_1}=6.7\times10^{-3}$. When the damage affects the middle cliques $\mathcal{C}_2$ and $\mathcal{C}_3$ the antifragile response is lower, with $\bar{\Lambda}_{\mathcal{C}_2}=5.1\times10^{-3}$ and $\bar{\Lambda}_{\mathcal{C}_3}=4.4\times10^{-3}$, respectively. It is worth noticing that the robustness exhibited when $\mathcal{C}_2$ or $\mathcal{C}_3$ are damaged is boosted in comparison to when the damage affects the entire structure. 
\\[2mm]
In Fig. \ref{Fig_4}(d), we show the results for a star of cliques. This structure is composed of eleven cliques (a central clique with each of its nodes connected to a pair of identical cliques). For this network, we identify two different responses to damage. On the one hand, it is observed antifragility when the damage affects the external cliques $\mathcal{C}_1$, as evidenced by $\bar{\Lambda}_{\mathcal{C}_1}=3.1\times10^{-3}$ and in the response $\langle\mathcal{F}(T)\rangle$. On the other hand, the central clique $\mathcal{C}_2$ displays fragility. For this clique  $\bar{\Lambda}_{\mathcal{C}_2}=-2.1\times10^{-3}$. The damage in this part of the network causes a degradation in the transport capacity of the structure, as indicated by $\langle\mathcal{F}(T)\rangle<1$ for different $T$. The response to damage in $\mathcal{C}_2$ is the most unfavorable of all, even compared to the case of damage distributed in all the links of the network.
\\[2mm]
In all the examples explored for networks with cliques, we find responses of antifragility, and in each case, the behavior of $\langle\mathcal{F}(T)\rangle$  aligns with the values obtained for $\bar{\Lambda}_{\mathcal{C}_i}$. Our findings for  $\bar{\Lambda}_{\mathcal{C}_i}$ show that this measure is a good indicator of antifragility since it is able to identify the communities that, when damaged, enhance the transport capacity of the networks. 
\subsubsection{Antifragility in networks with communities}
Our previous results reveal antifragility in networks with perfect communities (cliques). Let us now analyze the response to damage in networks with non-clique-like communities generated using Lancichinetti-Fortunato-Radicchi
(LFR) graphs \cite{PhysRevE.78.046110}. LFR graphs capture features of real-world networks, such as heterogeneity of degree distributions and community size distributions. We use the modularity $Q$ that for an undirected graph with $N_{\mathcal{C}}$ communities is given by
\begin{equation}\label{Q_def}
	Q\equiv \sum_{i=1}^{N_{\mathcal{C}}}\left[\frac{|\mathcal{E}_i|}{|\mathcal{E}|}-\left(\frac{k_{\mathcal{C}_i}}{2|\mathcal{E}|}\right)^2\right],
\end{equation}
where $k_{\mathcal{C}_i}$ is the sum of degrees of the nodes in the community $\mathcal{C}_i$.
The values of $Q$ in Eq. (\ref{Q_def}) classify the networks quantifying the accuracy of the division into communities by comparing the fraction of edges within communities with the expected value of the same quantity in a randomized network \cite{Newman_Girvan_2004,Clauset2004,Leicht2008}. We generate 240 LFR graphs with $N=100$ nodes and $0.238\leq Q\leq  0.742$ varying the mixing parameter $\mu$ (fraction of edges shared between community nodes and  other nodes in the network \cite{PhysRevE.78.046110}), from $\mu=0.05$ to $\mu=0.5$. Additionally,  we use degree distribution exponents of  2 and 3 and community size distribution exponents of 1 and 2 \cite{PhysRevE.78.046110}. For each network, $Q$ was calculated using the NetworkX (2.6.3) package. 
\\[2mm]
In Fig. \ref{Fig_5}, we present the results obtained for 240 LFR graphs. In order to illustrate the capability of these LFR graphs to offer an antifragile response to the damage within their communities,  in Fig. \ref{Fig_5}(a) we show
\begin{equation}
\Lambda^{(\mathrm{max})}_\mathcal{C}\equiv\max\left(\{\bar{\Lambda}_{\mathcal{C}_i}:i=1,\ldots, N_{\mathcal{C}}\}\right),
\end{equation}
i.e., the maximum value of $\bar{\Lambda}_{\mathcal{C}_i}$ for each graph, as a function of $Q$. Additionally,    the fraction $\nu$ of communities that trigger an antifragile response ($\bar{\Lambda}_{\mathcal{C}_i}>0$) in each graph is represented by the colorbar. We observe that in all the structures analyzed, there is at least one antifragile community, and the values of $\nu$ show that in most networks, the antifragile communities exceed $50\%$. Furthermore, the fractions $\nu$ do not show a specific dependence on $Q$, and we can observe networks with several antifragile communities even at low values of $Q$ within the analyzed range. Also, it is worth noticing that the values $\Lambda^{(\mathrm{max})}_\mathcal{C}$ increase with higher  values of $Q$. Additionally, we evaluate the effects of cumulative damage on four of these graphs. In Figs. \ref{Fig_5}(b)-(e), we present with continuous lines the values of $\langle\mathcal{F}(T)\rangle$ for networks with $Q=0.376,\,0.535,\, 0.653$, and $0.734$ when damage is located within communities. The thick dashed lines represent  $\langle\mathcal{F}(T)\rangle$ when damage spreads throughout the entire structure. The values $\langle\mathcal{F}(T)\rangle$ are calculated using  $10^4$ Monte Carlo simulations  with $\alpha=0.5$. The results show that each graph yields different responses to damage, depending on the community that is affected. The different values of $\bar{\Lambda}_{\mathcal{C}_i}$ for each community agree with this variety of responses.   In cases of communities with  $\bar{\Lambda}_{\mathcal{C}_i}\leq0$, we observe fragile responses with a rapid decrease of $\langle\mathcal{F}(T)\rangle$ or, at most, robust responses with a moderate decrease of $\langle\mathcal{F}(T)\rangle$. However,  the results show improvements in the transport capacity of the networks when damage is localized in communities with $\bar{\Lambda}_{\mathcal{C}_i}>0$. Particularly, the improvements, understood as the maximum value that $\langle\mathcal{F}(T)\rangle$ can achieve,  are significant for   $0.631\leq Q\leq  0.734$, corresponding to cases of well-defined communities. In both cases, the values of $\Lambda_{\mathcal{C}}^{(\mathrm{max})}$ are similar, but the maximum values of $\langle\mathcal{F}(T)\rangle$ are different. Our findings for the  LFR graphs explored suggest that the extent of the improvement depends on the topology of the network. Although not all densely connected communities are antifragile, such as the clique $\mathcal{C}_2$ in Fig. \ref{Fig_4}(d),  an antifragile community with a high value of $\bar{\Lambda}_{\mathcal{C}_i}$ and also densely connected may significantly enhance  $\langle\mathcal{F}(T)\rangle$ if subjected to damage.
\begin{figure*}[t!]
	\centering
	\includegraphics*[width=1.0\textwidth]{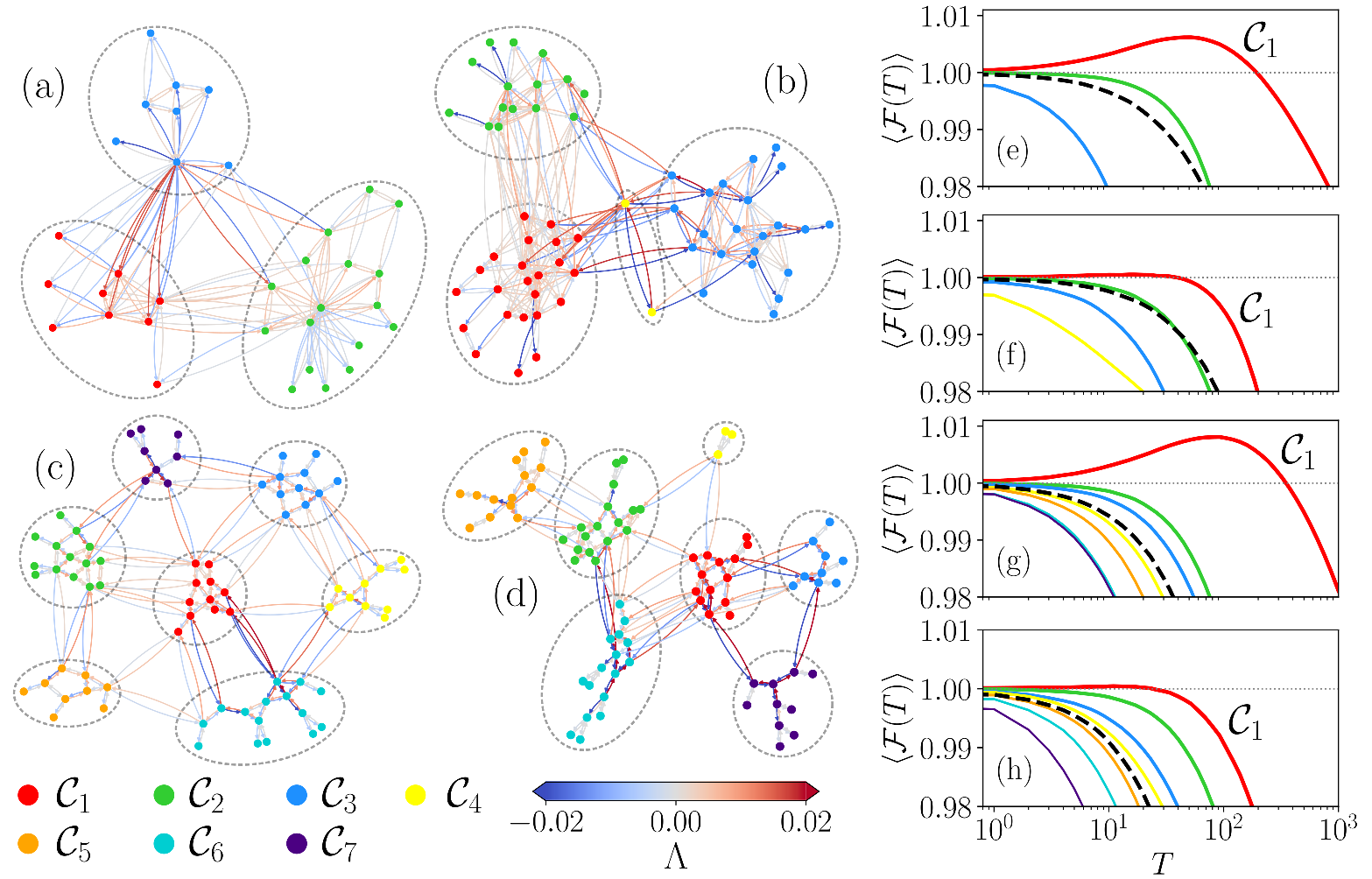}
	\vspace{-1mm}
	\caption{Response to damage in four real networks with communities. (a) The karate club network, (b) a dolphins network, and the metro systems in (c) Paris and (d) London. The nodes in each community $\mathcal{C}_i$ are presented with different colors codified at the bottom. The colors of the edges represent the values of $\Lambda$ obtained numerically using Eq. (\ref{Lambda_edge}) considering infinitesimal damage in each link. Panels (e)-(h) depict the respective $\langle \mathcal{F}(T)\rangle$ as a function of $T$ when damage is localized in the edges of a community (solid lines with the same colors as the nodes in the networks). The dashed curve corresponds to $\langle \mathcal{F}(T)\rangle$ when the damage affects the entire network and the horizontal dotted line represents $\langle \mathcal{F}(T)\rangle=1$. All the ensemble averages are calculated using $10^4$ Monte Carlo simulations with $\alpha=0.5$ (see details in the main text and Table \ref{Table_1}).}
	\label{Fig_6}
\end{figure*}
\begin{table}[h!]
	\begin{center}
		\begin{tabular}{c  c c c c c c c c c} 
			\hline 
			{\bf Network} & $N$ & $|\mathcal{E}|$ & $\tau(0)$  & $\mathcal{C}_{i}$ & $N_i$  & $\bar{\Lambda}_{\mathcal{C}_i}$  \\[1mm]
			\hline 
			\hline
			\multirow{3}{*}{Karate club \cite{Zachary1977}} & \multirow{3}{*}{34} & \multirow{3}{*}{78}  & \multirow{3}{*}{63.9} &  $\mathcal{C}_{1}$  &  9   &   0.001996 \\ 
            &    &    &              & $\mathcal{C}_{2}$  & 17   &   0.000608 \\
            &    &    &               & $\mathcal{C}_{3}$  & 8   &   -0.004638 \\
            \hline
			\multirow{4}{*}{Dolphins \cite{bottlenose}} & \multirow{4}{*}{62} & \multirow{4}{*}{159} &  \multirow{4}{*}{155.2} & $\mathcal{C}_{1}$  & 23   &   0.000757 \\
	        &    &    &               & $\mathcal{C}_{2}$  & 15   &   0.000288 \\
			&    &    &              & $\mathcal{C}_{3}$  & 22   &   -0.001077 \\
			&    &    &               & $\mathcal{C}_{4}$  & 2   &   -0.007719 \\
			\hline
			\multirow{7}{*}{Metro Paris \cite{SybilWolfram2014,Derrible2012}} & \multirow{7}{*}{78} & \multirow{7}{*}{125} & \multirow{7}{*}{239.2} & $\mathcal{C}_{1}$  & 11   &   0.002025 \\
			&    &    &              & $\mathcal{C}_{2}$  & 14   &   0.000927 \\
			&    &    &               & $\mathcal{C}_{3}$  & 11   &   0.000628 \\
			&    &    &               & $\mathcal{C}_{4}$  & 10   &   -0.000394 \\
			&    &    &              & $\mathcal{C}_{5}$  & 9   &   -0.001093 \\
			&    &    &               & $\mathcal{C}_{6}$  & 15   &   -0.002537 \\
			&    &    &               & $\mathcal{C}_{7}$  & 8   &   -0.002646 \\
				\hline
			\multirow{7}{*}{Metro London \cite{SybilWolfram2014,Derrible2012}} & \multirow{7}{*}{83} & \multirow{7}{*}{121} & \multirow{7}{*}{329.1} & $\mathcal{C}_{1}$  & 14   &   0.001797 \\
           &    &     &        & $\mathcal{C}_{2}$  & 17   &   0.00143 \\
           &    &     &        & $\mathcal{C}_{3}$  & 9   &   0.000427 \\
           &    &     &        & $\mathcal{C}_{4}$  & 3   &   -0.000229 \\
           &    &     &        & $\mathcal{C}_{5}$  & 14   &   -0.000476 \\
           &    &     &        & $\mathcal{C}_{6}$  & 16   &   -0.001433 \\
           &    &     &        & $\mathcal{C}_{7}$  & 10   &   -0.004369 \\
            \hline
		\end{tabular}
		\caption{\label{Table_1}  Characterization of the response to damage for diffusive transport in real networks with communities. We present the name of the network, the number of nodes $N$, the total number of edges $|\mathcal{E}|$, the global time $\tau(0)$ for the structure without damage, the different communities $\mathcal{C}_i$, the number of nodes $N_i$ in the community and the average $\bar{\Lambda}_{\mathcal{C}_i}$ (see details in the main text).} 
	\end{center}
\end{table}
\subsubsection{Real networks}
In this section, we explore the ideas introduced to measure antifragility in four real networks with communities. We study two well-known social networks: the karate club network ($N=34$) which shows the pattern of friendships between the members of a karate club at an American university in the 1970s \cite{Zachary1977}, and a dolphins network ($N=62$) describing the social interaction graph representing contact time between dolphins within a pod off the coast of New Zealand \cite{bottlenose}. In addition, we consider the metro networks in Paris ($N=78$) and London ($N=83$). In different urban transportation networks, the global function of the infrastructure is to communicate efficiently to all the nodes. In some cases, models with random walkers can help us to characterize the mobility in a particular transportation mode \cite{riascos2019,riascos2020} or to understand the effect of damage in the entire structure \cite{Eraso-metro}. The metro networks are obtained from Refs. \cite{SybilWolfram2014,Derrible2012}, in this case, the nodes represent stations where users can change between lines or the end stations of a line (see details in Ref. \cite{Derrible2012}). 
\\[2mm]
In Figs. \ref{Fig_6}(a)-(d), we depict the four real networks considered in our analysis. Each network is classified in communities $\mathcal{C}_i$, the nodes in each community are shown with different colors codified at the bottom of the panels. Also, in this representation, we plot each directed edge according to $\Lambda$  obtained numerically using Eq. (\ref{Lambda_edge}) considering infinitesimal damage localized in the specific link (the values are encoded in the colorbar). In Fig. \ref{Fig_6}(a) for the karate club, $Q=0.381$ obtained using Eq. (\ref{Q_def}) for a classification with three communities. Similarly, Fig. \ref{Fig_6}(b) shows the network of dolphins with $Q=0.495$ for four communities. For the metro of Paris in Fig. \ref{Fig_6}(c) $Q= 0.614$ and for London in Fig. \ref{Fig_6}(d) $Q=0.629$, both transportation networks have seven communities. 
\\[2mm]
On the other hand, in Figs. \ref{Fig_6}(e)-(h), we plot the curves  $\langle \mathcal{F}(T)\rangle$ as a function of $T$ for the implementation of the algorithm of preferential accumulation of damage in the edges. We use the value $\alpha=0.5$ and the averages are obtained with $10^4$ realizations. The different curves are obtained when damage affects only the links of a specific community $\mathcal{C}_i$. The colors used in the curves are the same of the nodes in the communities in the panels in Figs. \ref{Fig_6}(a)-(d). In addition, the dashed curve represents the case when damage affects all the links in the network [the results in Fig. \ref{Fig_6}(e) correspond to the karate club, Fig. \ref{Fig_6}(f) for the network of dolphins, Fig.  \ref{Fig_6}(g) for the metro in Paris and Fig. \ref{Fig_6}(h) for the metro in London].
\\[2mm]
In Table \ref{Table_1}, we report different values for the networks in Fig. \ref{Fig_6}, their community structure and their response to damage. We present the number of nodes $N$ and the number of links $|\mathcal{E}|$ in the undirected network (with  $2|\mathcal{E}|$ edges when it is necessary to consider the direction of the links). We also report the global time $\tau(0)$ defined in Eq. (\ref{tau_global0}) for the transport in the network without damage associated with the dynamics of random walkers following the standard random walk strategy. The values of  $\tau(0)$ give an estimate of the average number of steps needed by the random walker to reach a target from any initial condition \cite{ReviewJCN_2021}. In addition, we present the communities $\mathcal{C}_i$ in each network along with the respective number of nodes $N_i$ and the average $\bar{\Lambda}_{\mathcal{C}_i}$ obtained numerically using Eq. (\ref{Lambda_Community}). The communities in Fig. \ref{Fig_6} and in Table \ref{Table_1} are sorted in decreasing order of $\bar{\Lambda}_{\mathcal{C}_i}$.
\\[2mm]
The results in Figs.  \ref{Fig_6}(e)-(h) show how the response to damage observed in the curves $\langle \mathcal{F}(T)\rangle$ as a function of $T$ is well described by the average values  $\bar{\Lambda}_{\mathcal{C}_i}$ in Table \ref{Table_1}. In all the cases it is evidenced that at the level of communities a large $\bar{\Lambda}_{\mathcal{C}_i}$ implies a larger capacity to tolerate damage. Furthermore, the community $\mathcal{C}_1$ in the karate club [Fig.  \ref{Fig_6}(e)] and in the metro network of Paris [Fig.  \ref{Fig_6}(g)] presents an antifragile response with values $\langle \mathcal{F}(T)\rangle>1$ in the first part. Something similar occurs in the case of the metro network of London [Fig.  \ref{Fig_6}(h)]; however, although the values $\bar{\Lambda}_{\mathcal{C}_1}\approx 0.002$ are similar for these three networks, the antifragility of $\mathcal{C}_1$ observed through $\langle \mathcal{F}(T)\rangle$ is less pronounced in the case of London. Inquiring into the causes of this difference in the antifragile response of the $\mathcal{C}_1$ community for the two metro networks, it is observed that to understand in more detail the response to damage, it is useful to consider the standard deviation, denoted as $\sigma_{\mathcal{C}_i}$, for the values $\Lambda$ of the edges in the community. For example, for the karate network $\sigma_{\mathcal{C}_1}=0.0055$ and for the metro in Paris $\sigma_{\mathcal{C}_1}=0.0048$, whereas for the metro in London $\sigma_{\mathcal{C}_1}=0.011$ revealing that in this network, the community $\mathcal{C}_1$ has an antifragile response but has a larger dispersion of the values $\Lambda$ in its edges (the minimum value  $\Lambda_{\mathrm{min}}=-0.0245 $ and the maximum $\Lambda_{\mathrm{max}}=0.02678$), this particularity of  $\mathcal{C}_1$ in the metro of London explains the behavior observed in Fig.  \ref{Fig_6}(h).
\\[2mm]
All the analyses performed on the different networks explored in this research show that $\Lambda$ is an important quantity to examine the effect of damage in networks. This is a property of the edges of the network without damage and the dynamic process occurring in this structure. In the case of transport associated with random walkers, $\Lambda$ allows detecting antifragility at the level of edges. Using this information, it is also possible to characterize the damage response at the level of communities employing the average value $\bar{\Lambda}_{\mathcal{C}_i}$; however, this is a first approximation and it may be necessary to consider the standard deviation $\sigma_{\mathcal{C}_i}$ and even the whole distribution of the values of $\Lambda$ in the edges of each community.
\section{Conclusions}
\label{Sec_Conclusions}
In this paper, we explore the antifragility associated with the dynamics of random walks on networks. We use a functionality $\mathcal{F}_\beta$, defined in Eq. (\ref{F_beta_edge}), that measures the global capacity of transport when the weight of an edge is modified using $\beta$ and its derivative $\Lambda$ in Eq. (\ref{Lambda_edge}) evaluates the effect of modifications in the edges in the global performance of random walkers to explore a network. The value of $\Lambda$ can be seen as a property of each link, combining information on the connectivity and the dynamical process allowing to identify how the link responds to damage. The links with $\Lambda>0$ are antifragile, i.e., a moderate amount of damage on these connections increases the global performance of the system. Furthermore, the evaluation of the average $\bar{\Lambda}_{\mathcal{C}_i}$ in Eq. (\ref{Lambda_Community}) for the edges in a community $\mathcal{C}_i$ characterizes the effect of damage in densely connected groups of nodes. We tested the predictions of $\bar{\Lambda}_{\mathcal{C}_i}$ with the implementation of an algorithm of damage accumulation that reduces the capacity of the edges in the entire network or in specific communities.  Our findings show that heterogeneous networks with some densely connected groups of nodes may exhibit antifragility. In cases of  networks with well-defined communities, characterized by  high values of the modularity $Q$, we can find communities with large values of $\bar{\Lambda}_{\mathcal{C}_i}$ with a significant antifragile response. We also apply the formalism introduced to explore the response to damage in four real networks showing that in these structures $\bar{\Lambda}_{\mathcal{C}_i}$ characterizes the effect of damage accumulation at the level of communities. The approach developed in this research is general and can be applied to the study of other systems and processes throughout the definition of a global functionality and its response to changes in the interactions between the parts of the system. The methods introduced pave the way to a broad understanding of the emergence of antifragility and the impact of damage in complex systems.
\section*{Acknowledgment} 
LKEH acknowledges support from CONAHCYT México.
\section{Appendix}
\label{Appendix_Lambda}
In this Appendix, we explore analytically the effect of infinitesimal modifications of Markovian processes using perturbation theory. The main goal is to deduce an analytical expression for $\Lambda$ in Eq. (\ref{Lambda_def_Ts}) in terms of eigenvalues and eigenvectors of the transition probability matrix that defines the random walker in the original structure. Let us now apply perturbation theory to the eigenvalues and eigenvectors of  $\mathbf{\Pi}^{(a,b)}_\beta$ to determine $\Delta\lambda_\ell$, $\Delta P_j^\infty$, and $\Delta Z_{jj}^{(\ell)}$ and see their effects in $\Lambda$. 
We assume that the eigenvalues of the matrix $\mathbf{W}$ are nondegenerate. In this case, we can adapt the well-known methods of perturbation theory in quantum mechanics to the case of stochastic matrices. For example, at first-order approximation, the eigenvalues $\zeta_\ell$ of $\mathbf{\Pi}^{(a,b)}_\beta$ are
\begin{equation}
	\zeta_\ell=\lambda_\ell+\beta \langle\bar{\phi}_\ell| \mathbf{\Gamma} |\phi_\ell\rangle +o(\beta^2).
\end{equation}
Therefore
\begin{equation}\label{Delta_lambda}
	\Delta \lambda_\ell=\langle\bar{\phi}_\ell| \mathbf{\Gamma} |\phi_\ell\rangle. 
\end{equation}
In particular, for $\ell=1$, $\zeta_1=\lambda_1$ since 
\begin{align}\nonumber
	\Delta \lambda_1&=\langle\bar{\phi}_1| \mathbf{\Gamma} |\phi_1\rangle\\ \nonumber
	&=\sum_{l,m=1}^N \langle\bar{\phi}_1|l\rangle \mathbf{\Gamma}_{lm}\langle m |\phi_1\rangle=\sum_{l,m=1}^N P_l^\infty \mathbf{\Gamma}_{lm}\\
	&=\sum_{l=1}^N P_l^\infty \times \sum_{m=1}^N \mathbf{\Gamma}_{lm}=\sum_{l=1}^N P_l^\infty \times 0=0.
\end{align}
On the other hand, for the set of right eigenvectors $\left|\psi_\ell\right\rangle$ of $\mathbf{\Pi}^{(a,b)}_\beta$, perturbation theory requires
\begin{equation}\label{ReigVpsi}
	\left|\psi_\ell\right\rangle=
	\begin{cases}
		\displaystyle
		\left| \phi_1\right\rangle&\ell= 1,\\
		\displaystyle
		\left|\phi_\ell\right\rangle+\beta\sum_{m\neq \ell}\frac{|\phi_m \rangle \langle\bar{\phi}_m| \mathbf{\Gamma} |\phi_\ell\rangle }{\lambda_\ell-\lambda_m}  &\ell= 2,\ldots,N. \\
	\end{cases}
\end{equation}
In a similar manner, for the corresponding set of left eigenvectors $\left\langle \bar{\psi}_\ell\right|$ of $\mathbf{\Pi}^{(a,b)}_\beta$, it is necessary that the set of eigenvectors satisfy $\left\langle \bar{\psi}_m|\psi_\ell\right\rangle=\delta_{m\ell}$. The ansatz  $\left\langle \bar{\psi}_\ell\right|=\left\langle \bar{\phi}_\ell\right|+\beta\sum_{m\neq \ell} c_m \left\langle \bar{\phi}_m\right|$ leads to the set of eigenvectors
\begin{equation}\label{LeigVpsi}
	\left\langle\bar{\psi}_\ell\right|=
	\left\langle\bar{\phi}_\ell\right|
	+\beta\sum\limits_{m\neq \ell}^N\frac{\langle\bar{\phi}_\ell| \mathbf{\Gamma} |\phi_m\rangle}{\lambda_\ell-\lambda_m}\left\langle\bar{\phi}_m\right|.
\end{equation}
Both sets of eigenvectors in Eqs. (\ref{ReigVpsi}) and (\ref{LeigVpsi}) fulfill the conditions
\begin{align}
	\mathbf{\Pi}^{(a,b)}_\beta \left|\psi_\ell\right\rangle&=\zeta_\ell \left|\psi_\ell\right\rangle+o(\beta^2), \\
	\left\langle \bar{\psi}_\ell\right| \mathbf{\Pi}^{(a,b)}_\beta &=\zeta_\ell \left\langle \bar{\psi}_\ell\right| +o(\beta^2),
\end{align}    
and form an orthonormal base.
\\[2mm]
Once  the eigenvectors of the modified transition matrix are obtained, we can calculate the variations of  $Z_{jj}^{(\ell)}=\left\langle j|\phi_\ell\right\rangle \left\langle\bar{\phi}_\ell|j\right\rangle$ as follows
\begin{equation}
	\left\langle j|\psi_\ell\right\rangle\left\langle\bar{\psi}_\ell|j\right\rangle=Z_{jj}^{(\ell)}+\beta \Delta Z_{jj}^{(\ell)}+o(\beta^2),
\end{equation}
and, using Eqs. (\ref{ReigVpsi}) and (\ref{LeigVpsi}), we obtain for $\ell=1,2,\ldots,N$
\begin{multline}\label{DeltaZ_pert}
	\Delta Z_{jj}^{(\ell)}=\sum_{m\neq \ell}\frac{\left\langle j|\phi_m\right\rangle \left\langle\bar{\phi}_\ell|j\right\rangle\langle\bar{\phi}_m| \mathbf{\Gamma} |\phi_\ell\rangle}{\lambda_\ell-\lambda_m}\\
	+\sum_{m\neq \ell}\frac{\left\langle j|\phi_\ell\right\rangle \left\langle\bar{\phi}_m|j\right\rangle\langle\bar{\phi}_\ell| \mathbf{\Gamma} |\phi_m\rangle}{\lambda_\ell-\lambda_m}.
\end{multline}
For the particular case with $\ell=1$, using $\langle\bar{\phi}_m| \mathbf{\Gamma} |\phi_1\rangle=0$, we have
\begin{equation}\label{DeltaPinf}
	\Delta P_j^\infty=\Delta Z_{jj}^{(1)}=\sum_{m=2}^N\frac{ \left\langle\bar{\phi}_m|j\right\rangle \left\langle j|\phi_1\right\rangle \langle\bar{\phi}_1| \mathbf{\Gamma} |\phi_m\rangle}{1-\lambda_m}.
\end{equation}
Finally, introducing the results in Eqs. (\ref{Delta_lambda})-(\ref{DeltaPinf}) into the definitions of $\mathcal{T}_1$, $\mathcal{T}_2$, $\mathcal{T}_3$ in Eqs. (\ref{Time_1})-(\ref{Time_3}), we have
\begin{equation}
	\mathcal{T}_1=\sum_{\ell=2}^N \frac{1}{1-\lambda_\ell}X^{(\ell)},
\end{equation}
with
\begin{multline}
	X^{(\ell)}=\sum_{j=1}^N  \frac{1}{N P^\infty_j} \sum_{m\neq \ell}\frac{\left\langle j|\phi_m\right\rangle \left\langle\bar{\phi}_\ell|j\right\rangle\langle\bar{\phi}_m| \mathbf{\Gamma} |\phi_\ell\rangle}{\lambda_\ell-\lambda_m}
	\\
	+
	\sum_{j=1}^N  \frac{1}{N P^\infty_j}\sum_{m\neq \ell}
	\frac{
		\left\langle j|\phi_\ell\right\rangle \left\langle\bar{\phi}_m|j\right\rangle\langle\bar{\phi}_\ell| \mathbf{\Gamma} |\phi_m\rangle}{\lambda_\ell-\lambda_m},
\end{multline}
for $\ell=2,3,\ldots,N$. Similarly, 
\begin{align}
	\mathcal{T}_2&=\sum_{\ell=2}^N \frac{\langle\bar{\phi}_\ell| \mathbf{\Gamma} |\phi_\ell\rangle}{(1-\lambda_\ell)^2}\left(\sum_{j=1}^N  \frac{Z_{jj}^{(\ell)}}{N P^\infty_j} \right),\\
	\mathcal{T}_3&=\sum_{j=1}^N \frac{\tau_j(0)}{N P^\infty_j}\left(\sum_{m=2}^N\frac{ \left\langle\bar{\phi}_m|j\right\rangle \left\langle j|\phi_1\right\rangle \langle\bar{\phi}_1| \mathbf{\Gamma} |\phi_m\rangle}{1-\lambda_m}\right).
\end{align}	

\onecolumngrid

\providecommand{\noopsort}[1]{}\providecommand{\singleletter}[1]{#1}%
\end{document}